\documentclass[9pt,a4paper,twocolumn]{article}
\input{pisika.dat}
\usepackage{amsmath,amssymb} 
\usepackage{subfigure}

\begin{document}

\providecommand{\ShortAuthorList}[0]{}
\title{Learning Arbitrary Complex Matrices by Interlacing Amplitude and Phase Masks with Fixed Unitary Operations}
\author[1,2,$\dagger$]{Matthew Markowitz}
\author[1,$\dagger$]{Kevin Zelaya}
\author[1,2,*,$\dagger$]{Mohammad-Ali Miri}
\affil[1]{Department of Physics, Queens College of the City University of New York, Queens, New York 11367, USA}
\affil[2]{Physics Program, The Graduate Center, City University of New York, New York, New York 10016, USA}
\affil[*]{Corresponding author: mmiri@qc.cuny.edu}
\affil[$\dagger$]{These authors contributed equally to this work}

\begin{abstract}
\noindent

Programmable photonic integrated circuits represent an emerging technology that amalgamates photonics and electronics, paving the way for light-based information processing at high speeds and low power consumption. Considering their wide range of applications as one of the most fundamental mathematical operations there has been a particular interest in programmable photonic circuits that perform matrix-vector multiplication. In this regard, there has been great interest in developing novel circuit architectures for performing matrix operations that are compatible with the existing photonic integrated circuit technology which can thus be reliably implemented. Recently, it has been shown that discrete linear unitary operations can be parameterized through diagonal phase parameters interlaced with a fixed operator that enables efficient photonic realization of unitary operations by cascading phase shifter arrays interlaced with a multiport component. Here, we show that such a decomposition is only a special case of a much broader class of factorizations that allow for parametrizing arbitrary complex matrices in terms of diagonal matrices alternating with a fixed unitary matrix. Thus, we introduce a novel architecture for physically implementing discrete linear operations. The proposed architecture is built on representing an $N \times N$ matrix operator in terms of $N+1$ amplitude-and-phase modulation layers interlaced with a fixed unitary layer that could be implemented via a coupled waveguide array. The proposed architecture enables the development of novel families of programmable photonic circuits for on-chip analog information processing.

\DOI{}
\end{abstract}

\maketitle
\thispagestyle{titlestyle}

\section{Introduction}

\begin{figure*}[ht]
\centering 
\includegraphics[width=0.85\textwidth]{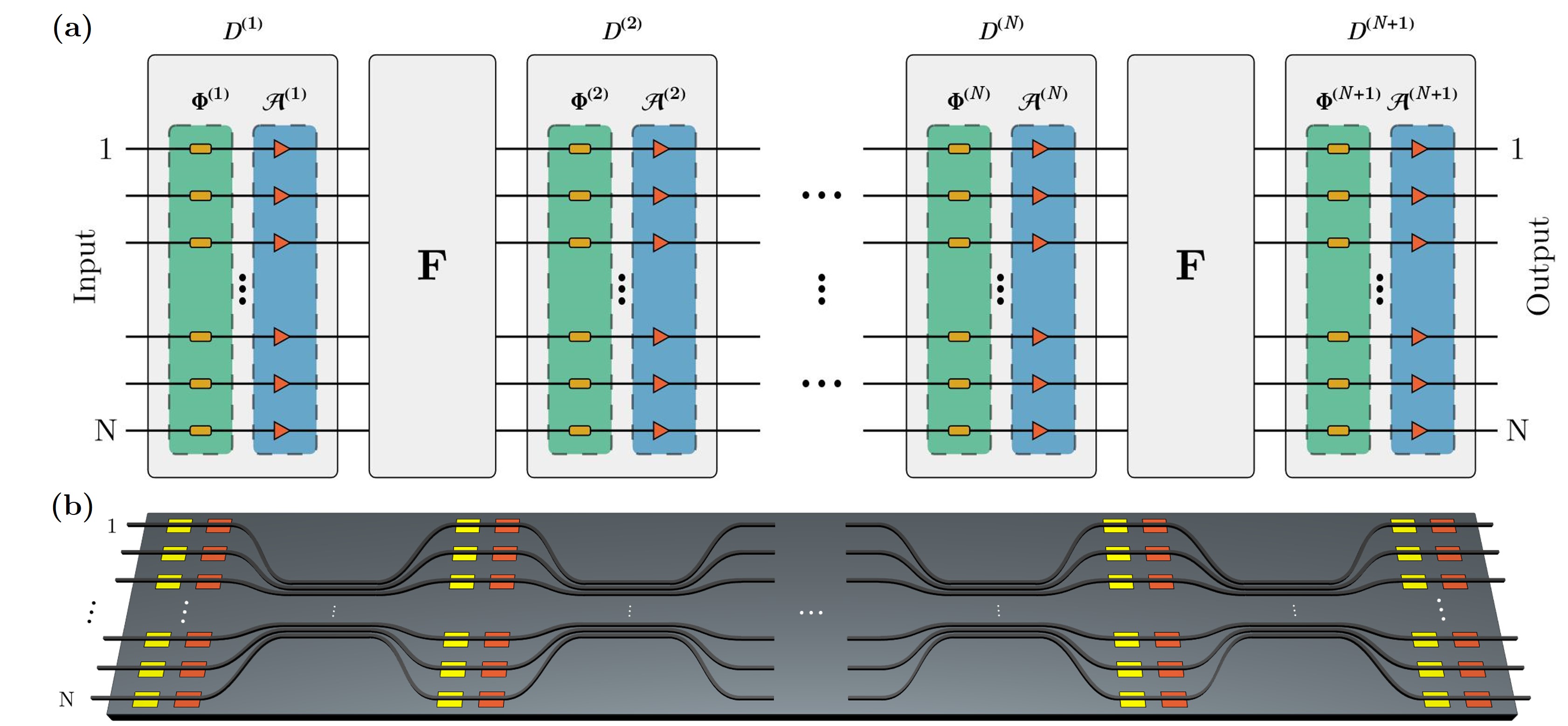}
\caption{(a) A block diagram sketching the proposed factorization in~\eqref{univ_A}, with $D^{(m)}$ the complex diagonal matrices and $F$ the DFrFT matrix acting as the unitary interlacing matrix. Furthermore, $D^{(m)}$ is decomposed as the product of a layer of phase shifters (orange rectangles) $\Phi^{(m)}$ and a layer of amplitude modulators (red triangles) $\mathcal{A}^{(m)}$. (b) A schematic of a photonic device architecture with waveguide couplers representing the intervening operator $F$ and with amplitude (red) and phase (yellow) modulators representing the programmable diagonal matrices.}
\label{fig1}
\end{figure*}

The prospect of realizing reconfigurable photonic circuits that perform arbitrary discrete linear operations on light has gained significant interest in the past decade \cite{harris2018linear,bogaerts2020programmable}. This is driven by the wide range of tasks that can be characterized by matrix operations, subsequently leading to the development of a chip-scale platform to deploy general-purpose light-based applications in classical and quantum information processing~\cite{notaros2017programmable,harris2017quantum,slussarenko2019photonic}. An efficient programmable photonic matrix-vector multiplier opens numerous opportunities such as enabling on-chip photonic neural networks \cite{shen2017deep,shastri2021photonics,wetzstein2020inference}, novel photonic interconnects \cite{miller2010optical,Christos2018interconnect}, and multistate quantum processors \cite{,carine2020multi,madsen2022quantum,lu2020three}. In the context of integrated photonics, previous efforts have been focused on realizing circuits that perform linear unitary operations, a subclass of matrix operations that preserve the norms. This is driven primarily by the interesting properties of unitary groups and the rigorous mathematical parameterization and factorization of unitary matrices in terms of simpler unitary elements with specific optical implementation. In particular, it has been shown that all unitaries can be realized with meshes of Mach-Zehnder interferometers incorporating phase shifters in different circuit configurations~\cite{reck1994experimental, miller2013self,clements2016optimal,Shokraneh2020,rahbardar2023addressing,yu2023heavy}. Architectures based on mode divisor multiplexing have proved to be another handy resource~\cite{tang_integrated_2017,ling2023chip}. On the other hand, recently, an alternative factorization of unitaries has been suggested that involves interlacing diagonal phase modulations with a fixed intervening unitary operator \cite{tanomura_robust_2020,pastor2021arbitrary,saygin_robust_2020, Skryabin2021, tanomura2023multi, Markowitz23a, markowitz2023auto, Zelaya_2023_random-F}. In particular, we have previously shown that by interlacing nonuniform waveguide couplers that emulate the Discrete Fractional Fourier Transform (DFrFT) with $N+1$ layers of phase shifters one can represent arbitrary unitary operations \cite{Markowitz23a}. The robustness of this architecture versus imperfections and defects makes it suitable for implementation \cite{markowitz2023auto}. Furthermore, we have shown that the fixed intervening operator can be extended to a wide range of operations that subsequently allow for different physical realizations of the proposed programmable device architecture \cite{Zelaya_2023_random-F}.

Despite numerous efforts to realize unitary operators, to the best of our knowledge, no prior work discusses a direct implementation of general non-unitary matrices. The latter is particularly fascinating, as it allows for general matrix-vector multiplications not possible using only unitaries. Even though the singular value decomposition (SVD) facilitates the creation of these matrices, through the combination of two unitaries sandwiching a positive semidefinite diagonal matrix, it inevitably results in devices with large optical path lengths. On the other hand,  no direct compact factorization of arbitrary complex non-unitary matrices exists. In most relevant prior works, it has been shown that every complex $2^n \times 2^n$ matrix can be factored as the finite product of circulant and diagonal matrices~\cite{Muller1998algorithmic, Schmid2000decomposing}. 
Furthermore, Huhtanen and Per\"am\"aki~\cite{Huhtanen2015factoring} have demonstrated that any $N\times N$ complex-valued matrix can be decomposed as the product of no more than $2N-1$ interlaced circulant and diagonal matrices. This result has been particularly useful in deep-learning models for video classification~\cite{araujo2018training}, optical networking~\cite{Lukens2020}, and frequency encoding in quantum information processing~\cite{lukens2017frequency}. 

In this work, we introduce a layer-efficient factorization to represent arbitrary complex-valued matrices. This is achieved after modifying the factorization presented in~\cite{Huhtanen2015factoring} by limiting the maximum number of interlaced layers and replacing the interlaced DFT layer with a DFrFT. The latter is inspired by photonic applications, where platforms to deploy on-chip DFrFT based on waveguide arrays are known in the literature~\cite{weimann2016implementation}, as well as other optical elements to represent the factorization discussed here. Numerical experiments based on optimization algorithms reveal that such a truncation leads to numerical error as low as that obtained using the exact factorization. The case of unitary matrices is discussed as a particular limit, where the number of parameters to be optimized reduces to one-half of the number required for the general problem. Remarkably, the architecture is shown to be robust against random perturbations on the DFrFT waveguide, provided that the latter are within acceptable tolerance errors. Rigorous electromagnetic simulations further corroborate the realization of the proposed programmable photonic device.

\section{Results}
\subsection{Model}
It has been known that an arbitrary complex-valued $N\times N$ matrix $A$ is parameterized in terms of at most $2N-1$ interlaced circulant and diagonal matrices. In turn, a circulant matrix ($C^{(m)}$) can be diagonalized through the discrete Fourier transform (DFT) and the inverse discrete Fourier transform (IDFT) matrices, $\mathbb{F}$ and $\mathbb{F}^{-1}$, according to the relation $C^{(m)}=\mathbb{F} D^{(m)} \mathbb{F}^{-1}$ \cite{Gray2006toeplitz}. Therefore, an arbitrary complex matrix $A$ can be parameterized through complex diagonal matrices interlacing with the DFT matrix and its inverse as $A = D^{(2N-1)} \mathbb{F} D^{(2N-2)} \mathbb{F}^{-1} D^{(2N-3)} \cdots D^{(3)} \mathbb{F} D^{(2)} \mathbb{F}^{-1} D^{(1)}$. Since a $N \times N$ diagonal matrix contains $N$ complex elements, the latter factorization involves $4N^2-2N$ real parameters in total. This relation defines a vastly over-parameterized problem containing redundant factoring layers given that $2N^2$ real parameters are required to characterize any $N\times N$ complex-valued matrix. 

Here, we consider and analyze the modified factorization
\begin{equation}
\label{univ_A}
    A = D^{(M)} F D^{(M-1)} F D^{(M-2)} \cdots D^{(3)} F D^{(2)} F D^{(1)},
\end{equation}
where $F$ is a unitary matrix to be defined and $D^{(m)}$ represent diagonal matrices with components $D^{(m)}_{p,q}=d^{(m)}_{p}e^{i\phi^{(m)}_{p}}\delta_{p,q}$, for $m\in\{1, \ldots, M\}$. In the latter, $\phi_{p}^{(m)}\in(-\pi,\pi]$ and $d_{p}^{(m)}\geq 0$ represent the $p$-th phase and amplitude parameters of the $m$-th complex-valued diagonal layer. In this context, Fig.~\ref{fig1}(a) illustrates a block diagram of the proposed matrix decomposition and the corresponding photonic architecture is sketched in Fig.~\ref{fig1}. Here, the unitary matrix $F$ is realized using an array of coupled waveguides, while each diagonal matrix is composed of two contiguous layers, one containing exclusively $N$ phase modulators ($\Phi^{(m)}$) next to another layer comprised of $N$ amplitude modulators ($\mathcal{A}^{(m)}$).

In the proposed factorization, the $M$ complex-valued diagonal matrices $D^{(m)}$ require $2MN$ real parameters. We numerically explore this ansatz and analyze its universality, i.e., the capability of the factorization~\eqref{univ_A} to reconstruct all arbitrary $N\times N$ complex-valued matrices. The exact number of complex diagonal matrices $M$ that guarantees universality is also determined numerically. We focus on a unitary matrix $F$ related to the discrete fractional Fourier transform. In this regard, it is worth remarking that several valid definitions for the discrete fractional Fourier transform exist in the literature~\cite{Ata97,Can00}. Here, we adapt the definition from Ref.~\cite{weimann2016implementation}, based on the wave evolution of guided modes through a non-uniformly spaced optical lattice, henceforth called Jx lattice. This is particularly handy since its physical realization is straightforward \cite{weimann2016implementation, HonariLatifpour2022, Keshavarz2023}. In this fashion, the DFrFT matrix of fractional order $\alpha=\pi/2$ writes as the propagator operator for the normalized length $\pi/2$ as $F = \textup{exp}(i H \pi/2 )$, where, $H$ is the $J_x$ lattice Hamiltonian with matrix components $H_{p,q}=\kappa_{p}\delta_{p+1,q}+\kappa_{p-1}\delta_{p-1,q}$ and hopping rates $\kappa_{p} = \frac{1}{2} \sqrt{(N-p)p}$, for $p\in\{ 1,\ldots,N-1\}$ ~\cite{weimann2016implementation}. This coupling is physically achieved by using identical single-mode photonic waveguides with non-uniform spacing \cite{weimann2016implementation, HonariLatifpour2022}. In the microwave domain, such a coupling has been shown possible using parallel microstrip lines coupled to interdigital capcitors~\cite{Keshavarz2023}. It is straightforward to show that $F$ satisfies all properties of the discrete fractional Fourier transform, e.g., $F$ is unitary, $F^2$ is the parity operator, and $F^4$ is the identity matrix. Furthermore, the DFrFT defined by $F$ converges to the DFT in the limit of large $N$. 

\subsection{Optmization and Universality}
To demonstrate the universality of this device, we optimize the respective phases $\phi_{p}^{(m)}$ and amplitudes $d_{p}^{(m)}$, where $p\in\{1,\ldots, N\}$ and $m\in\{1,\ldots, M\}$, for an ensemble of randomly chosen target complex-valued matrices $A_t$. The latter are generated using the singular value decomposition~\cite{Moon2000SignalProcessing} (SVD) so that each target matrix decomposes as $A_{t}=U\Sigma V^{\dagger}$, where $U$ and $V^{\dagger}$ are unitary matrices randomly generated in accordance with the Haar measure \cite{mezzadri_how_2007}. In turn, the diagonal matrix $\Sigma=diag(\sigma_{1},\ldots,\sigma_{N})$ contains the singular values of $A_{t}$, with $\sigma_{i}\geq 0$ randomly taken from the interval $[0.25,1]$. The goodness of approximation of the target matrices is explored against the number $M$ of complex diagonal matrices, which can, in turn, be understood as composed of $M$ phase layers and $M$ amplitude layers, corresponding to $2NM$ phase parameters. The loss function, interchangeably called error norm, is defined as the mean square error
\begin{equation}
\label{eq_L}
L = \frac{1}{N^2} \| A - A_t \|^2,
\end{equation}
where $\|A\|=\sqrt{\text{Tr}(A^{\dagger}A)}$ is the Frobenius norm, $A_{t}$ is the complex-valued target matrix under consideration, and $A$ is the reconstructed matrix using~\eqref{univ_A}. 

The error norm is a multivariate function defined on the parameter space $\{\phi_{p}^{(m)},d_{p}^{(m)}\}_{p=1,m=1}^{N,M}$. The latter can either be convex or non-convex in nature, the global minimum location of which defines the solution as long as it vanishes. Numerically, the error norm never vanishes, and the global minimum might not be unique. Thus, to find an acceptable solution, we define the error norm tolerance value $L_c$ so that a (local) minimum is accepted as a solution whenever $L<L_{c}$.

The optimization is performed using the Levenberg-Marquardt algorithm (LMA), which is well suited to sum-of-squares objective functions and can be used for both under and over-determined problems \cite{levenberg1944method}. The universality of the factorization~\eqref{univ_A} is tested by randomly generating 100 target matrices for $N=4$ and $N=6$ according to the SVD decomposition previously discussed. The parameter space is randomly initialized by assigning the values $\phi_{p}^{(m)}\in(-\pi,\pi]$ and $d_{p}^{(m)}\in[0,\ell]$, for $p\in\{1,\ldots, N\}$ and $m\in\{1,\ldots, M\}$, where $\ell$ sets the upper bound for the search region of the optimized amplitudes $d_{p}^{(m)}$. Our numerical runs indeed show that the computational time scales up with higher values of $\ell$. The LMA is run to find the parameters that lead to an error norm below a prefixed tolerance $L_{c}$ while truncating the amplitude parameters to $\ell \leq 1.5$ in most cases, which typically yields satisfactory results. However, in some instances, $\ell$ may be increased slightly to enhance the optimization performance.

\begin{figure}[ht]
\centering 
\includegraphics[width=0.4\textwidth]{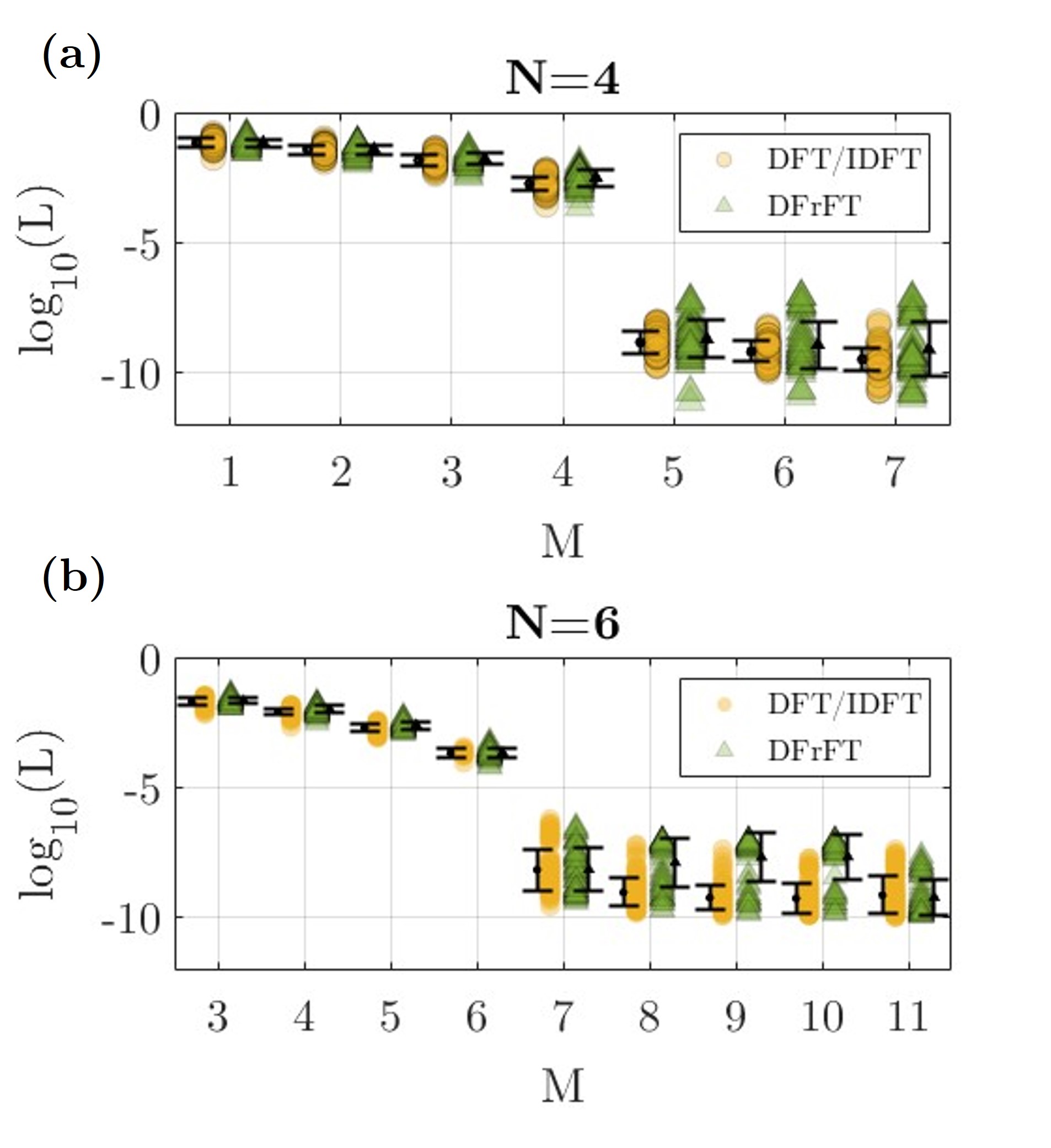}
\caption{Error norms $L$ for $N=4$ (a) and $N=6$ (b) while varying the the number of layers $M$. The 100 targets were used for each case, and the LMA is run until the error tolerance $L_c=10^{-7}$ is reached. This is performed using two different choices of the intervening operator $F$ interlaced with diagonal layers. The circle markers correspond to using alternating DFT and IDFT layers, while the triangle markers correspond to using DFrFT. The upper bound on the magnitude of the diagonal amplitudes $d_{i}^{(m)}$ was set to 1.5.}
\label{fig2}
\end{figure}

The latter procedure is carried out for matrices of fixed dimension $N$ while varying the number of layers $M$. Furthermore, to establish a reference point to compare our findings, we consider the factorization based on interlaced layers of DFT and IDFT operations proposed in Ref.~\cite{Huhtanen2015factoring} truncated to $M$ layers (where $M\leq 2N-1$). By doing this, we aim to recover the exact factorization for $M=2N-1$ while analyzing its behavior for fewer layers. Simultaneously, the factorization based on the DFrFT in~\eqref{univ_A} is also considered for the same number of layers. This establishes a benchmark scenario where performance increases can be monitored as a function of $M$ for different $N$. The resulting error norms associated with the optimization process are shown in Fig.~\ref{fig2}(a) and Fig.~\ref{fig2}(b) for $N=4$ and $N=6$, respectively. From both figures, the exact factorization at $M=2N-1$, using the interlaced DFT and IDFT factorization, reveals an error norm around $10^{-7}$. Such a result shows the numerical precision expected during the optimization process, from which the error tolerance is established at $L_{c}=10^{-7}$ to allow for small deviations. For all other values of $M$, the error norm is monitored for each configuration without imposing any tolerance value $L_c$. In both cases, the error norm shows mild improvements at each stage while spanning the number of layers from $M=N-3$ to $M=N$. Still, these errors are above the required tolerance $L_{c}$. Notably, between $M=N$ and $M=N+1$ layers, the error norm undergoes a phase transition, where it drastically drops from $L\sim 10^{-4}$ to $L \sim 10^{-7}$, the desired tolerance. Mild improvements in the error norm are observed for $M=N+2,\ldots,2N-1$, which are not relatively significant in terms of performance. This verifies that~\eqref{univ_A} produces results numerically equivalent to the exact case discussed in Ref.~\cite{Huhtanen2015factoring}. We thus fix 
\begin{equation}
    M=N+1
\end{equation}
henceforth as the effective number of layers required for the factorization~\eqref{univ_A} throughout the manuscript.

\begin{figure*}[h!]
\centering
\includegraphics[width=0.75\textwidth]{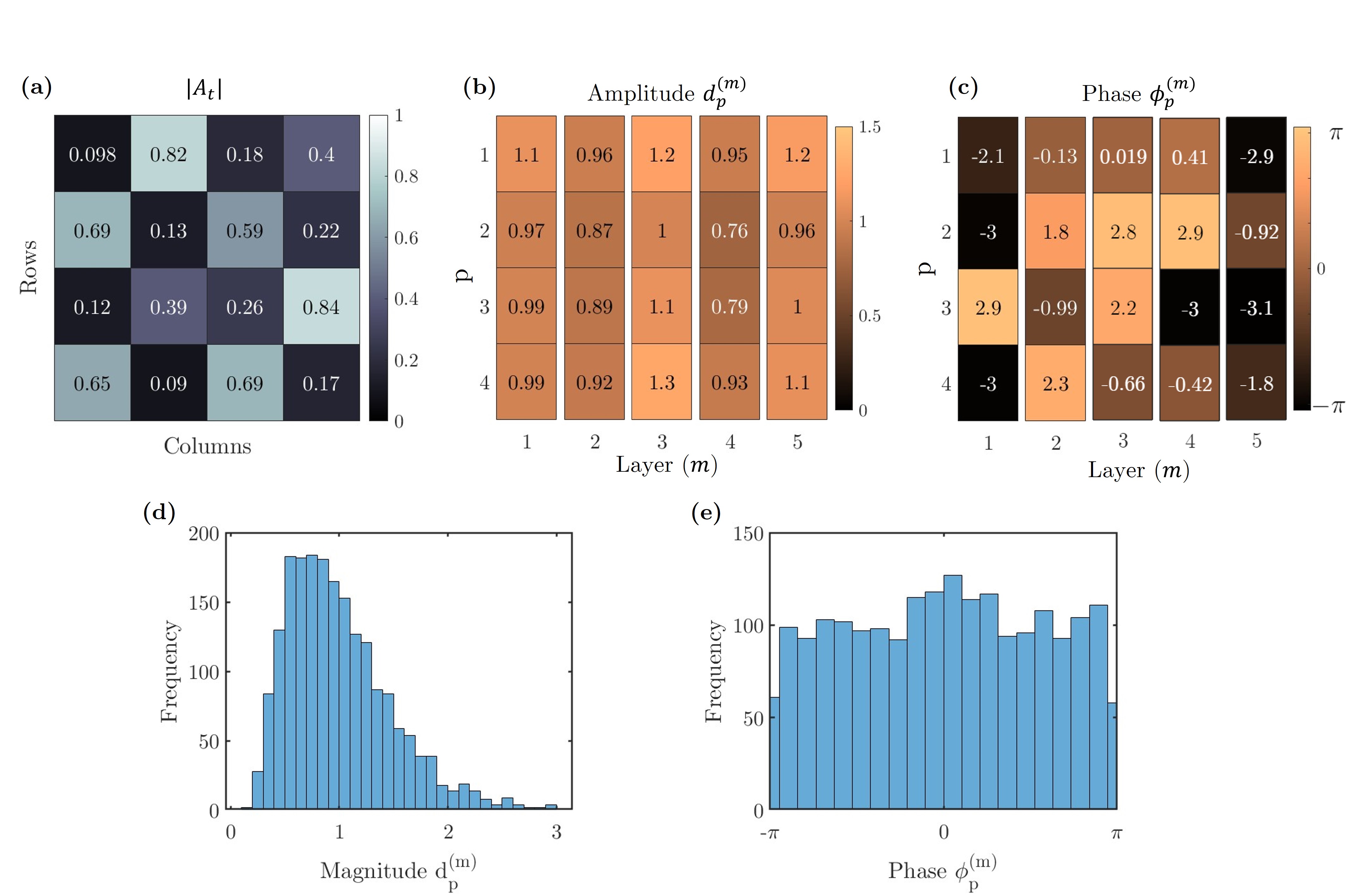}
\caption{(a) Example of a random target matrix for $N=4$ and $M=5$ generated using the SVD with two-Haar random unitary matrices and singular values uniformly chosen in the range [0.25,1]. Optimized amplitude (b) and phase (c) parameters for the target matrix shown in (a). Distribution of the magnitudes (d) and phases (e) across all target matrices for $N=4$ and $M=5$.}
\label{fig3}
\end{figure*}

To further illustrate the nature of the optimized parameters, let us consider $N=4$ and $M=N+1=5$ as illustrative examples. Here, we consider a set of 100 random matrices generated with singular values uniformly distributed so that $\sigma_{j}\in[0.25,1]$, with $j\in\{1,2,3,4\}$, for each target matrix, whereas  $U$ and $V^{\dagger}$ are Haar random matrices as customary. In this form, we generate random complex-valued matrices with specific eigenvalues, and we can study the maximum bound $\ell$ for the amplitude parameters $d_{p}^{(m)}$ to be optimized. Fig.~\ref{fig3}(a) displays one example from the set of target matrices, where the corresponding optimized magnitudes and phases are illustrated in Fig.~\ref{fig3}(b) and Fig.~\ref{fig3}(c), respectively. In turn, Fig.~\ref{fig3}(d) shows the occurrence distribution frequency of the optimized amplitude parameters $d_{p}^{(m)}$ for all the 100 target matrices. The latter reveals that the values of $d_{p}^{(m)}$ are more likely to fall within the range $(0,2)$ when the singular values of the target matrix are less than one. Although there may exist some $d_{p}^{(m)}>2$, their probability of occurrence is relatively low. Thus, one can focus mainly on the previous interval during optimization and increase the interval upper bound $\ell$ for tuning purposes when required. This is particularly useful since the number of parameters increases quadratically with $N$, and the optimization becomes computationally demanding. On the other hand, amplitude parameters in the interval $(1,2)$ appear with considerable frequency, implying that active gain elements are required in the physical realization unless the desired matrix is scaled with a global amplitude factor, which is discussed below. The latter behavior is not present for the optimization of the phase parameters $\phi_{p}^{(m)}$, as these are constrained to the compact domain $(-\pi,\pi]$. Fig.~\ref{fig3}(d) shows indeed that the optimized phases do not follow any particular pattern during the optimization of random targets, and their occurrence frequency distributes almost uniformly across $(-\pi,\pi]$.

\subsection{Unitary Limit}
\begin{figure}[h]
\centering
\includegraphics[width=0.4\textwidth]{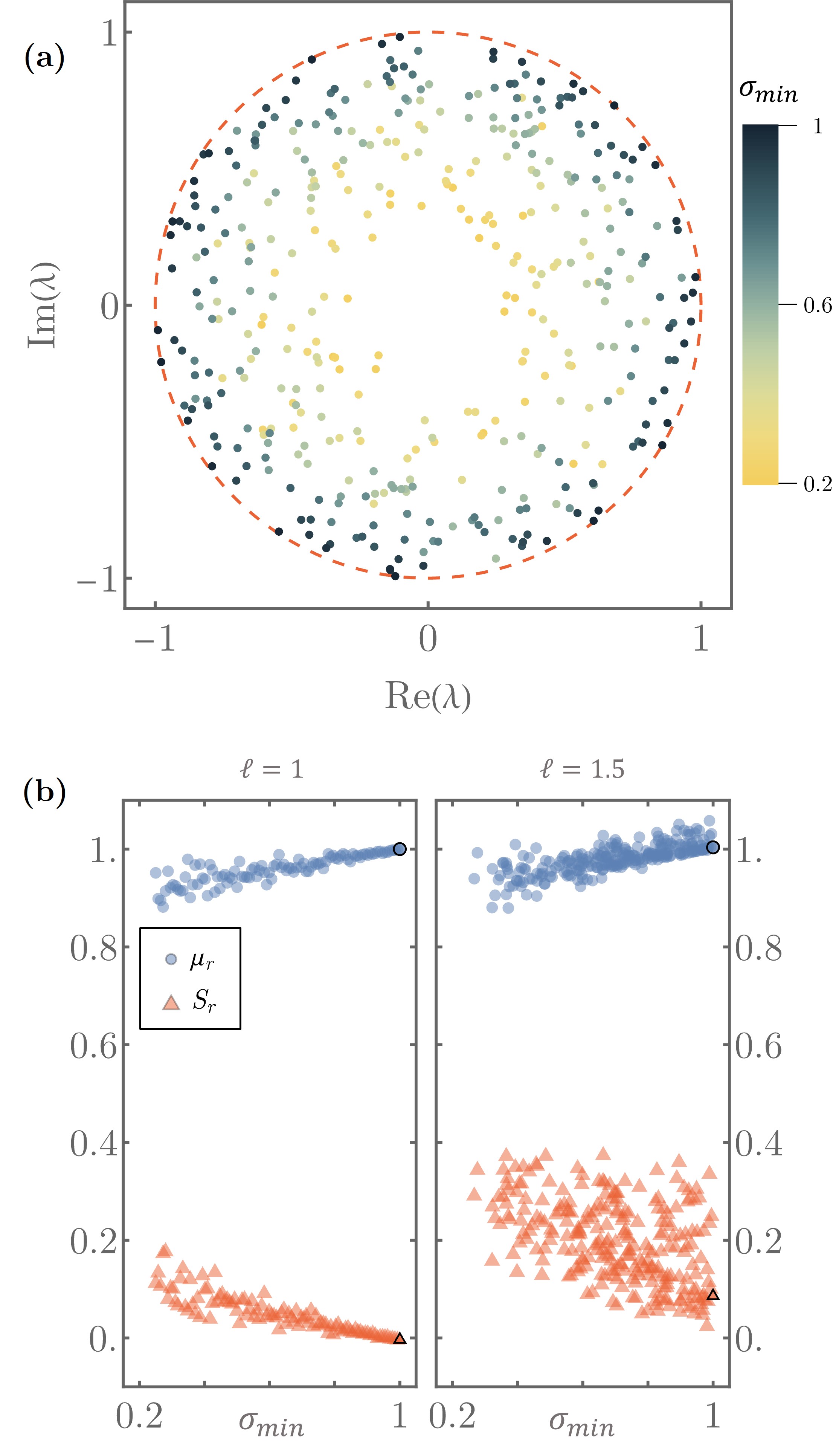}
\caption{(a) Eigenvalues $\lambda$ of 100 Haar randomly generated $4\times 4$ target matrices with singular values $\sigma_{1,2,3,4}\in[\sigma_{min},1]$ for increasing values of $\sigma_{min}$. (b) Mean $\mu_{r}$ and standard deviation $S_{r}$ of the set of optimized amplitude parameters $r=\{d_{p}^{(m)}\}_{p,m=1}^{N,M}$, for N=4, M=5, and optimization search region limited by $\ell$, as a function of $\sigma_{min}$. Edged markers correspond to the unitary case ($\sigma_{min}=1$).}
\label{fig4}
\end{figure}
\noindent
Unitary target matrices $A_{t}^{u}\in U(N)$ are a subset contained in the set of complex-valued $N \times N$ matrices, and their factorization and photonic implementation have been widely explored in the literature~\cite{zhou_tunable_2018,tanomura_robust_2020,saygin_robust_2020,pastor2021arbitrary,Markowitz23a}. It is a natural question to explore whether our architecture shall reproduce the results for unitaries as a special case. To address this, one can factor the diagonal matrices of \eqref{univ_A} as $D^{(m)}=\Phi^{(m)}\mathcal{A}^{(m)}$, with $\Phi^{(m)}$ and $\mathcal{A}^{(m)}$ diagonal matrices with components $\Phi^{(m)}_{p,q}=e^{i\phi^{(m)}_{p}}\delta_{p,q}$ and $\mathcal{A}^{(m)}_{p,q}=d^{(m)}_{p}\delta_{p,q}$, where in particular $\Phi^{(m)}$ is unitary. In general, the product of unitary matrices renders a unitary matrix. Following the factorization~\eqref{univ_A} and $D^{(m)}=\Phi^{(m)}\mathcal{A}^{(m)}$, it is thus straightforward that a unitary target matrix is obtained from the factorization if $\mathcal{A}^{(m)}=\mathbb{I}$ for all $m\in\{1,\ldots, N\}$, with $\mathbb{I}$ the $N\times N$ identity matrix. In turn, the singular values of $A_{t}^{u}$ are $\sigma_{p}=1$ for all $p\in\{1,\ldots,N\}$. 

The latter is a sufficient condition, yet not necessary, as there might be cases in which the optimized amplitudes $d_{p}^{(m)}\neq 1$ render unitary matrices. We thus numerically explore this possibility. The behavior around the vicinity of unitary matrices can be studied by first randomly constructing complex-valued matrices whose singular values $\sigma_{p}\leq 1$. By using the convention $\sigma_{p}\geq\sigma_{p+1}$ for all $p\in\{1,\ldots,N-1\}$, it is then enough to make $\sigma_{min}:=min(\{\sigma_{p}\}_{p=1}^{N})=\sigma_{N}\approx 1$ to ensure that $A_{t}$ approximates to a unitary matrix $A_{t}^{u}$. The unitary limit of $A_{t}=U\Sigma V^{\dagger}$ is independent of the random unitary Haar matrices $U$ and $V^{\dagger}$ and depends exclusively on the choice of the singular values. This is illustrated by considering $N=4$ and randomly generating 100 target matrices so that $\sigma_{min}=\sigma_{4}$ spans uniformly through the interval $[0.2,1]$ for each target. The remaining singular values $\sigma_{j=1,2,3}$ are randomly distributed in the interval $[\sigma_{min},1]$ for each selection of $\sigma_{min}$. The eigenvalues $\lambda$ for each of the resulting target matrices are shown in Fig.~\ref{fig4}(a), which distribute from the inner part of the complex unit circle $\vert\lambda\vert<1$ ($\sigma_{min}<1$) and converge to the unit circle in the unitary case limit ($\sigma_{min}\rightarrow 1$). The corresponding optimization is performed for each of the random target matrices, from which the set $r=\{d_{p}^{(m)}\}_{p,m=1}^{N, N+1}$ is constructed for each target so that it contains the optimized amplitude parameters exclusively.

Indeed, during the optimization process, the phases are automatically determined, but we are particularly interested in the amplitudes, as they provide the relevant information around the unitary limit. The mean $\mu_{r}$ and standard deviation $S_{r}$ of $r$ for each target are illustrated in Fig.~\ref{fig4}(b). The latter shows how the optimized amplitudes distribute when $\sigma_{min}$ increases from 0.2 to the unity for two optimization scenarios. In the first case, shown in Figure~\ref{fig4}(b) (left panel), the optimization region where the values of $d_{p}^{(m)}$ are searched is bound from above to $\ell=1$. Here, no particular pattern is observed for small $\sigma_{min}\approx 0.2$, which is expected as the random complex-valued targets have no particular structure. In turn, a quite characteristic tendency is observed for $\sigma_{min}\rightarrow 1$, where the standard deviation tends to vanish, implying that the optimized amplitude parameters start to pile up around the vicinity of $\mu_{r}=1$. Although this numerical evidence suggests that unitaries are recovered when all the optimized amplitudes approach one, the parameter search has been performed in the interval $0\leq d_{p}^{(m)}\leq \ell=1$. Thus, the possibility of finding unitaries for $d_{p}^{(m)}>\ell$ shall not be ruled out. This is indeed the second case depicted in Figure~\ref{fig4}(b) (right panel), where the search region has been expanded, $0\leq d_{p}^{(m)}\leq\ell=1.5$. The transition to the unitary case is fuzzier and, although $\mu_r\rightarrow 1$, the standard deviation $S_r$ does not vanish; e.g., the amplitude parameters $d_{p}^{(m)}$ are strictly different from unity. To explain this, suppose that $N-1$ amplitude layers are equal to the identity and one is proportional to the identity, namely, $\mathcal{A}^{(m_1)}=d_{m_1}\mathbb{I}$ for $\lambda_1<1$ and some $m_{1}\in\{1,\ldots,N+1\}$. Thus, a second diagonal amplitude layer $\mathcal{A}^{(m_{2})}$ could exist so that $\mathcal{A}^{(m_{2})}=d_{m_1}^{-1}\mathbb{I}$, leading to a unitary matrix $A$. Although both cases render the required unitary matrix, the former is preferable as it is less computationally expensive.


\subsection{Defects and Error Mitigation }
\begin{figure}[h]
\centering
\includegraphics[width=0.45\textwidth]{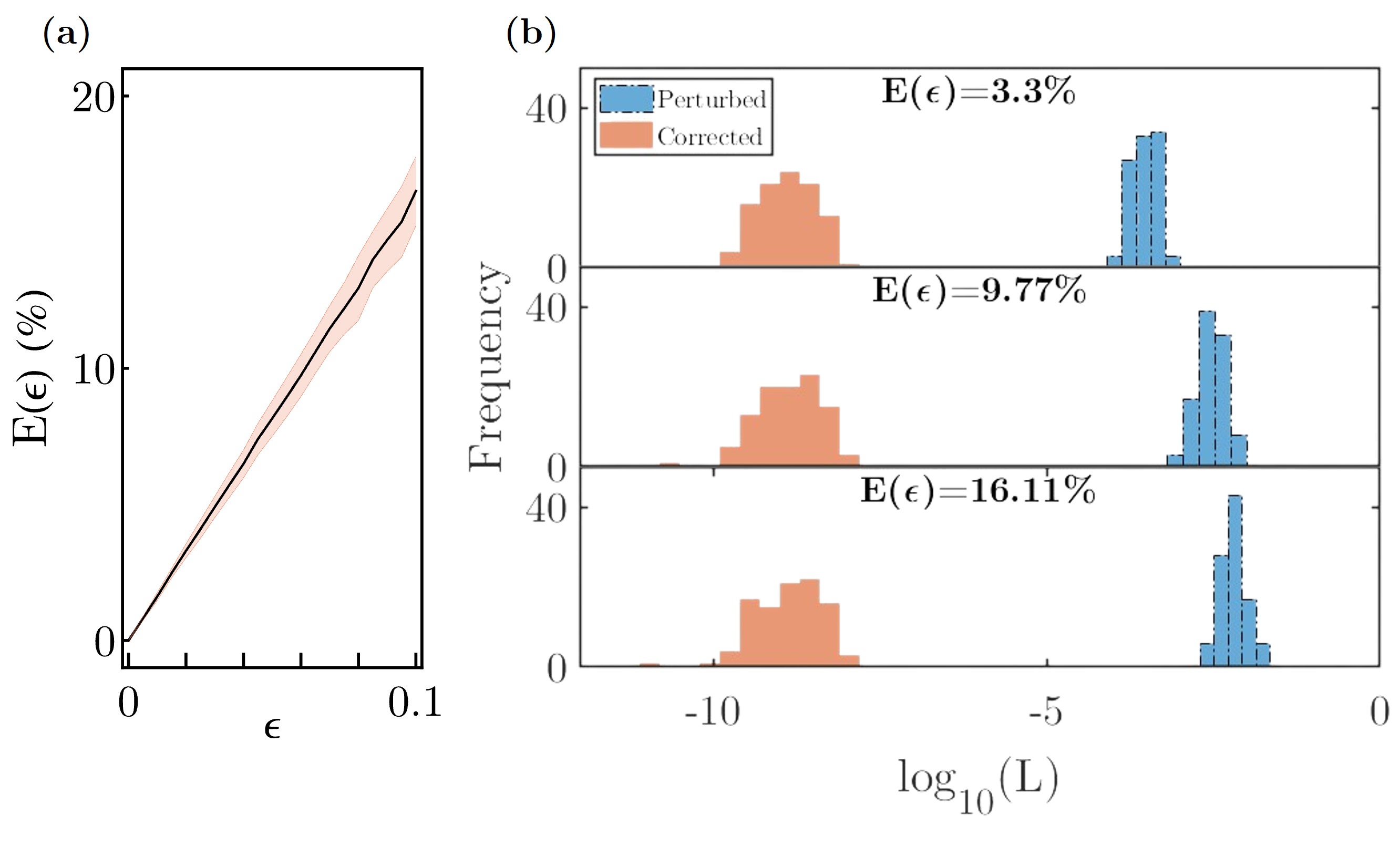}
\caption{(a) Mean and standard deviations of the percentage error $E(\epsilon)$ computed from 100 random target matrices per value of $\epsilon$. (b) Calibration when the mixing layers $F$ are perturbed: $\widetilde{F} = F+ \epsilon R$ where $R$ is a matrix with its elements chosen randomly within the complex unit circle. Shown are log-norms when the original amplitude and phase parameters are used (blue) and when re-calibrating the amplitude and phase parameters (orange).}
\label{fig5}
\end{figure}

So far, the universality of the architecture described by Eq.~\eqref{univ_A} has been verified. Nevertheless, one can consider a less idealistic scenario in which the interlaced DFrFT unitary matrix $F$ contains imperfections due to fabrication errors. Previous unitary factorizations~\cite{markowitz2023auto} have been shown to be resilient to perturbance on the interlacing matrix $F$ when the perturbation is considered to be unitary as well. This was achieved by perturbing the Hamiltonian defining the unitary evolution $F = \textup{exp}(i H \pi/2 )$. Here, we account for error in a general manner by considering the perturbation $\widetilde{F}(\epsilon)=F+\epsilon R$, where $\epsilon\ll 1$ is a perturbation strength parameter and $R$ a complex-valued random matrix whose components are constrained as $\vert R_{p,q}\vert<1$. This ensures that the new, non-unitary, interlacing matrix $\widetilde{F}$ does not significantly deviate from the DFrFT for small enough $\epsilon$. The value of $\epsilon$ by itself does not provide insight into the error introduced to the system. We thus estimate deviations from the ideal model through the relative percentage error $E(\epsilon):=\left(\Vert \widetilde{F}(\epsilon)-F\Vert/\Vert F \Vert\right) \times 100\% \equiv \epsilon \left( \Vert R\Vert/ \Vert F \Vert \right) \times 100\%$, which grows linearly with $\epsilon$. To visualize the error induced into the matrix $\widetilde{F}$, we consider 100 random matrices $R$ for each value of $\epsilon$ and compute the corresponding percentage error $E(\epsilon)$. Fig.~\ref{fig5}(a) displays the mean (solid line) and standard deviation (shaded area) for the set of points generated for each $\epsilon$. Particularly, one can see that perturbations induce errors about $3.3\%$, $9.77\%$, and $16.11\%$ for $\epsilon=0.02$, $\epsilon=0.06$, and $\epsilon=0.1$, respectively. This reveals that, although $\epsilon=0.1$ may be considered a small perturbation, it accounts for a considerable error in $\widetilde{F}(\epsilon)$. 

From the previous considerations, we test the effects of $\widetilde{F}$ in the interlaced architecture. To this end, a set of 100 Haar random and complex-valued target matrices are constructed for each parameter strength $\epsilon=0.02,0.06,0.10$. Likewise, a complex-valued random matrix $R$ is generated for each target matrix. First, the set of phase and amplitude parameters $\{\phi_{p}^{(n)},d_{p}^{(n)}\}_{p=1,n=1}^{N,N+1}$ are optimized when $\epsilon=0$ (unperturbed), which are next used to evaluate the interlaced architecture with $\widetilde{F}(\epsilon)$. The resulting error norms (blue bars) are presented in Fig.~\ref{fig5}(b), where it is clear that in all instances, the error norm is always higher than $10^{-5}$ and above the fixed tolerance $L_{c}$, even for relatively small deviations of $3.3\%$ ($\epsilon=0.02$). Despite this issue, it is possible to perform a second optimization on the interlaced architecture using the perturbed matrix $\widetilde{F}(\epsilon)$ and the previously found set $\{\phi_{p}^{(n)},d_{p}^{(n)}\}_{p=1,n=1}^{N,N+1}$ as the initialization parameters in the LMA minimization search. This leads to the new optimized parameters $\{\widetilde{\phi}_{p}^{(n)},\widetilde{d}_{p}^{(n)}\}_{p=1,n=1}^{N,N+1}$. The updated error norm shown in Fig.~\ref{fig5}(b) (orange bars) reveals that, by performing the second optimization, the error norm drops back under the tolerance value $L_{c}$. Therefore, the architecture~\eqref{univ_A} is resilient to random defects on the DFrFT layer, which can be amended by properly tuning the parameters. Indeed, this holds whenever defects lie within reasonable perturbation strengths; otherwise, the resulting interlaced matrix $\widetilde{F}(\epsilon)$ becomes mostly a random matrix.
%
\subsection{Photonic Device Design and Simulation}
\begin{figure*}[h]
    \centering
    \includegraphics[width=0.95\textwidth]{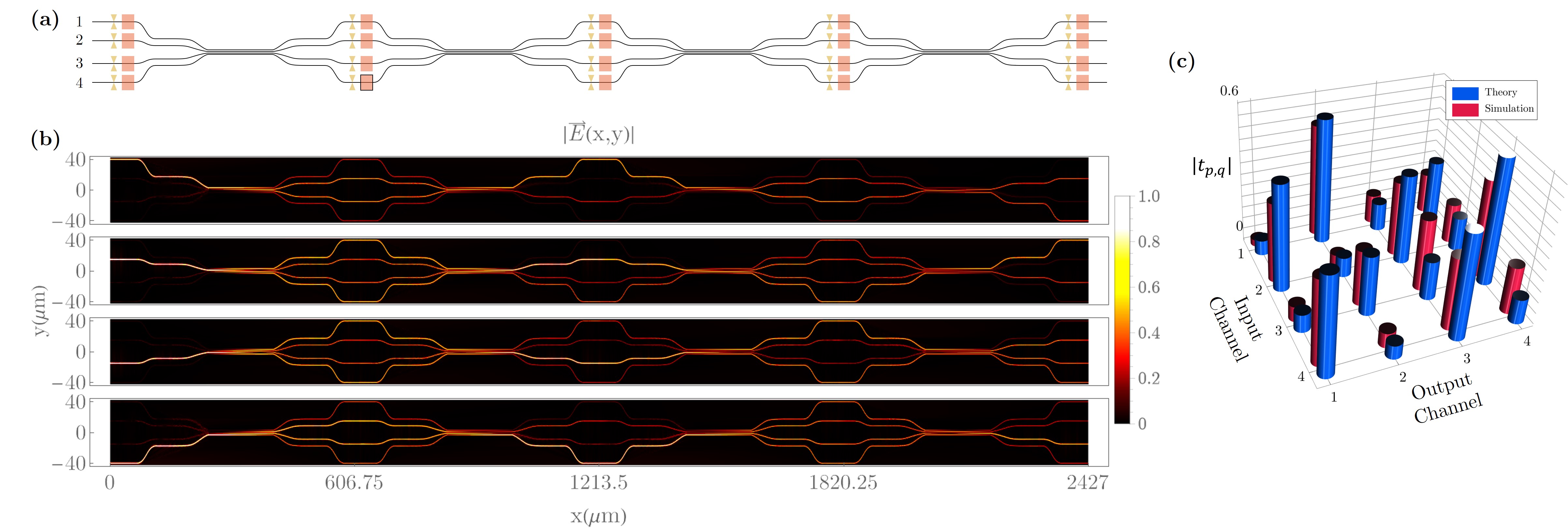}
    \caption{(a) A sketch of the proposed architecture for $N=4$ based on coupled waveguide arrays and amplitude (red) and phase (yellow) modulation using phase-change materials (PCMs). (b) The electric field $\vert \vec{E}(x,y) \vert$ intensity plot for the complex-matrix factorization~\eqref{univ_A} obtained from wave simulations when the phase and amplitude parameters are those depicted in Figs.~\ref{fig3}(b)-(c). The simulations are performed for the transverse electric field polarization (TE). (c) Theoretical and simulated amplitudes of the field transmission matrix elements $|t_{pq}|$, where the transmission matrix elements are defined as $t_{pq}=b_{p}/a_{q}$, i.e., the ratio of the electric field complex modal amplitude measured at the output channel $p$ over the electric field complex modal amplitude at the input channel $q$. The rescaled version of the parameters in Fig.~\ref{fig3}(b)-(c) has been used in both theory and simulation.}
    \label{fig6}
\end{figure*}
To further verify the proposed architecture, we perform rigorous wave propagation simulations using the finite elements method for a four-channel device ($N=4$) designed to operate at the telecommunication wavelength ($1550$ nm). Here, we consider silicon-on-insulator (SOI) ridge waveguides of width $500$nm and height $200$nm, while we consider refractive indices of $n_{c}=3.47$ and $n_{s}=1.4711$ for the core and substrate regions respectively. Using the effective index method, one can effectively approximate the problem into a highly accurate two-dimensional model, where the core is replaced by an effective core index $n_{eff}=2.7241$ for the propagation of the fundamental TE mode. In this case, we consider wave propagating along the $x$-direction and spatially varying along the $y$-direction, leading to an electric field of the form $\vec{E}=\mathcal{E}(x,y)\hat{z}$. The Jx lattice in this example comprises four waveguides with a separation of 1.5 $\mu m$ in the inner-most channels and 1.56132 $\mu m$ for the outer-most channels. This ensures that the DFrFT of order  $\pi/2$ is achieved for a total propagation length of 158 $\mu m$. The phase shifters are simulated by changing the refractive index of the waveguide to achieve the desired change. These are $20\,\mu m\times 20\,\mu m$ components (yellow-bowties in Fig.~\ref{fig6}(a)), based on the specifications of phase-change materials (PCM)~\cite{Rios2022ultra,youngblood2023integrated}. In turn, the amplitude modulators (red-rectangles in Fig.~\ref{fig6}(a)) are simulated through optical media with varying imaginary parts, a property available using PCM materials such as Ge2Sb2Te~\cite{zhang2019broadband}, which allow for compact and relatively robust extinction rate due to strong deviations on the imaginary part of the refractive index. 

An illustration of the propagation of the electric field amplitude related to the equivalent two-dimensional structure is presented in Fig.~\ref{fig6}(b) when the architecture is independently excited at each input. For testing purposes, the phase shifters and amplitude parameters have been loaded with the values shown in Fig.~\ref{fig3}(b) and Fig.~\ref{fig3}(c), respectively. Note that some amplitude parameters $d_{p}^{(m)}$ are larger than one, and one thus requires intensity gain in the proposed architecture, which is not being considered here. It is thus more convenient to reconstruct the rescaled target matrix $\widetilde{A}_{t}=\mathcal{D}^{-1}A_{t}$, with $\mathcal{D}=\prod_{m=1}^{N}\textnormal{max}(\{d^{(m)}_{q}\}_{q=1}^{N})$, which requires the amplitude parameters $\widetilde{d}_{p}^{(m)}=d_{p}^{(m)}/\textnormal{max}(\{d_{q}^{(m)}\}_{q=1}^{N})$ for its reconstruction. In the latter, we have factored out the largest $d_{p}^{(m)}$ per each layer $m$. By considering $\widetilde{A}_{t}$, we can eliminate the need for amplitude gain elements in the architecture, simplifying the design and reducing the manipulation of amplitude parameters. Specifically, one limits the manipulation of amplitude parameters to $N-1$ per layer instead of scaling every parameter, as the initially largest per layer is now fixed to one while the rest are rescaled accordingly. 

For comparison, the theoretical and simulated amplitudes of the field transmission matrix $t_{p,q}=b_p/a_q$ are shown in Fig.~\ref{fig6}(c). In this notation, $b_p$ represents the complex modal amplitude of the electric field in the output channel $p$, and $a_q$ represents the complex modal amplitude of the electric field in the input channel $q$. Here, the excitation is swept across all the $p$ input channels while the output is measured at the $q$-th channel. These results show some mild deviations of the simulation from the theory, possibly due to several factors, such as imperfect coupling in the waveguide array, amplitude modulators, and phase shifters. Despite this, the mean-square error $L$ of the simulated output with respect to the theoretical rescaled target $\mathcal{A}_{t}$ leads to $L=6.62918\times 10^{-3}$. The latter error can be improved by first identifying the defects on the lattice and accordingly modifying the interlacing layer $F$; then, the parameters shall be optimized a second time so that the defects of $F$ are taken into account. This is indeed the procedure discussed in~\cite{markowitz2023auto} for unitary architectures, which can be adapted for the present device.
\section{Discussion and Conclusion}
In summary, we proposed a novel architecture for implementing arbitrary discrete linear operators with photonic integrated circuits. The proposed architecture is built on interlacing fixed discrete fractional Fourier transform layers with programmable amplitude-and-phase modulator layers. Our results indicate that this architecture can universally represent $N \times N$ complex-valued operations with at most $N+1$ controllable layers. The proposed architecture offers a fairly simple physical realization by utilizing photonic waveguide arrays in conjunction with arrays of amplitude and phase modulators. It should be noted that even if the singular values of the target matrix are less than one, in general optimizing the amplitude modulators requires both gains ($d_{p}^{(m)}>1$) and losses ($d_{p}^{(m)}<1$). To overcome this challenge and to bypass the demanding requirement for gain, the reconstruction of the properly rescaled target matrix $\mathcal{A}_{t}$. Although modulators based on Mach-Zehnder interferometers or optical ring resonators~\cite{xu2006cascaded} are known optical implementations for amplitude modulators, such elements substantially increase the overall length and complexity of the final device, leading to potentially additional errors due to the presence of the coupling elements. Alternatively, phase-change materials based on Ge2Sb2Te and related structures~\cite{zhang2019broadband,youngblood2023integrated} are fascinating emerging candidates to replace and implement the required amplitude modulation. In the microwave domain, integrated active elements that produce both amplitude gains and losses are relatively simple to deploy and implement into the architecture. Additionally, a microwave realization of the DFrFT using microstrip and interdigital capacitors has been demonstrated and experimentally verified~\cite{Keshavarz2023}.

Potential manufacturing errors on the waveguide array characterizing the DFrFT have been taken into account and shown overall resilience on the universality performance of the architecture, provided that such errors lie within relative tolerance errors. Particularly, perturbed $F$ matrices with a percentage error of up $18\%$ have been considered. Although the optimized parameters obtained from the unperturbed lattice produce a significant error norm $L$ when defects are taken into account, a second optimization brings the error back to the imposed tolerance error. Therefore, defects can be mitigated overall by adequately tuning the available passive elements in the architecture.

Furthermore, well-known factorization schemes based on unitary targets are recovered as a particular case of the current architecture. Indeed, this is the case when all the amplitude modulators are set to unity, as the resulting factorization becomes the product of only unitary matrices. This can be numerically explored by studying the convergence of the amplitude parameters when all the singular values of the target matrix approach the unity. In such a case, it is found that the optimized amplitude parameters start piling up around the vicinity of one, which reduces the optimization space from $2N(N+1))$ real parameters to $N(N+1)$ phase elements. It is worth mentioning that the latter was performed by bounding the parameter search space in order to speed up the optimization process, and thus, other cases may exist where unitary matrices are recovered.\\

\noindent
\textbf{Funding}. This project is supported by the U.S. Air Force Office of Scientific Research (AFOSR) Young Investigator Program (YIP) Award FA9550-22-1-0189.\\

\noindent
\textbf{Disclosures}. Patent Pending

\bibliographystyle{pisikabst}
\bibliography{bibfile}

\end{document}